\documentclass[aps,preprint,superscriptaddress]{revtex4}
\usepackage{graphicx}
\usepackage{amsmath}
\usepackage{dcolumn}
\usepackage{color}
\usepackage{epstopdf}

\def\V{\textbf{V}}
\def\v{\textbf{v}}

\def\B{\textbf{B}}

\def\e{\textbf{e}}
\def\pa{\partial}

\begin{document}

\title{Resistive and ferritic-wall plasma dynamos in a sphere}

\author{I. V. Khalzov}
\affiliation{University of Wisconsin-Madison, 1150 University Avenue, Madison, Wisconsin 53706, USA}
\affiliation{Center for Magnetic Self Organization in Laboratory and Astrophysical Plasmas}
\author{B. P. Brown}
\affiliation{University of Wisconsin-Madison, 1150 University Avenue, Madison, Wisconsin 53706, USA}
\affiliation{Center for Magnetic Self Organization in Laboratory and Astrophysical Plasmas}
\author{E. J. Kaplan}
\affiliation{University of Wisconsin-Madison, 1150 University Avenue, Madison, Wisconsin 53706, USA}
\affiliation{Center for Magnetic Self Organization in Laboratory and Astrophysical Plasmas}
\author{N. Katz}
\affiliation{University of Wisconsin-Madison, 1150 University Avenue, Madison, Wisconsin 53706, USA}
\affiliation{Center for Magnetic Self Organization in Laboratory and Astrophysical Plasmas}
\author{C.~Paz-Soldan}
\affiliation{University of Wisconsin-Madison, 1150 University Avenue, Madison, Wisconsin 53706, USA}
\affiliation{Center for Magnetic Self Organization in Laboratory and Astrophysical Plasmas}
\author{K. Rahbarnia}
\affiliation{University of Wisconsin-Madison, 1150 University Avenue, Madison, Wisconsin 53706, USA}
\affiliation{Center for Magnetic Self Organization in Laboratory and Astrophysical Plasmas}
\author{E. J. Spence}
\affiliation{Princeton Plasma Physics Laboratory, P.O. Box 451, Princeton, New Jersey 08543, USA}
\affiliation{Center for Magnetic Self Organization in Laboratory and Astrophysical Plasmas}
\author{C. B. Forest}
\affiliation{University of Wisconsin-Madison, 1150 University Avenue, Madison, Wisconsin 53706, USA}
\affiliation{Center for Magnetic Self Organization in Laboratory and Astrophysical Plasmas}

\date{\today}

\begin{abstract}
We numerically study  the effects of varying electric conductivity and magnetic permeability of the bounding wall on a kinematic dynamo in a sphere for parameters relevant to Madison plasma dynamo experiment (MPDX). The dynamo is excited by a laminar, axisymmetric flow of von K\'arm\'an type. The flow is  obtained as a solution to the Navier-Stokes equation for an isothermal fluid with a velocity profile specified at the sphere's boundary. The properties of the wall are taken into account as thin-wall boundary conditions imposed on the magnetic field. It is found that an increase in the permeability of the wall reduces the critical magnetic Reynolds number $Rm_{cr}$. An increase in the conductivity of the wall leaves $Rm_{cr}$ unaffected, but reduces the dynamo growth rate.
\end{abstract}

\maketitle

Over the past decade, significant effort has been directed at the experimental demonstration of dynamo action --  self-excitation and maintenance of the magnetic field in a flowing electrically conducting fluid. A number of experiments with liquid metals have been constructed to test this phenomenon in various settings \cite{Peffley_2000, Forest_2002, Bourgoin_2002, Gailitis_2000, Stieglitz_2001, Monchaux_2007}, and successful observations of dynamo action have been reported in three of them \cite{Gailitis_2000, Stieglitz_2001, Monchaux_2007}. Among other things, these experiments revealed the critical importance of the magnetic properties of the flow-driving impellers. Namely, the von K\'arm\'an sodium experiment only self-sustained a dynamo field  if the impellers were ferromagnetic \cite{Monchaux_2007, Verhille_2010}. In addition, the finite resistivity of the experimental  container is expected to be crucial for the dynamo instability -- a situation similar to the resistive wall mode (RWM) in tokamaks \cite{Chu_2010}. Normally stable for the perfectly conducting wall, the RWM  can become unstable if the wall has finite resistivity; in this case the instability develops on the wall's resistive time scale.  These facts initiated more thorough theoretical studies of the effect of the imposed boundary conditions on the dynamo in experimentally relevant models \cite{Avalos_2003, Avalos_2005, Laguerre_2006, Gissinger_2008, Gissinger_2009, Roberts_2010, Guervilly_2010, Giesecke_2010}. The studies established that there are no general dependences of dynamo properties on  conductivity and permeability of the boundary; the dependences are different for different models and flows. 

\begin{figure}[bt]
\centering
\includegraphics[scale=1]{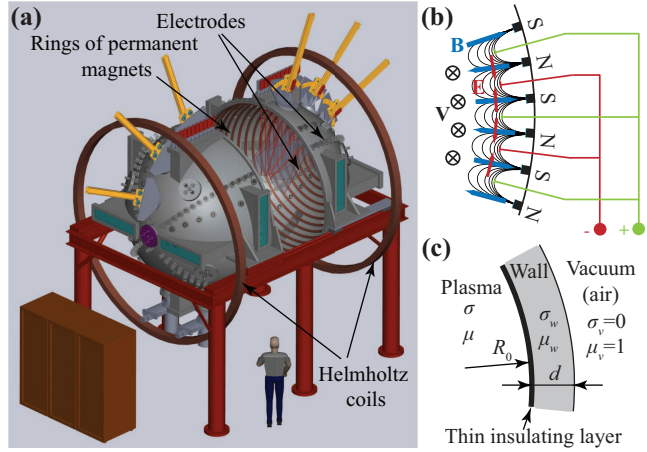}
\caption{Madison plasma dynamo experiment (MPDX): (a) sketch of the experiment; (b) electrode configuration near the wall for driving plasma velocity $v_\phi(\theta)$; (c) model for deriving thin-wall boundary conditions, with $\sigma$ and $\mu$ denoting conductivity and relative permeability of the respective media.}\label{MPDX_fig}
\end{figure}

This circumstance motivates us to perform an analogous study for the Madison plasma dynamo experiment (MPDX, Fig.~\ref{MPDX_fig}), currently under construction at the University of Wisconsin-Madison. The experiment is aimed at investigations of fundamental properties of dynamos excited by controllable flows of plasmas. Its original design was proposed in Refs.~\cite{Forest_2008, Spence_2009} and  conceptual features were successfully tested in the plasma Couette experiment (PCX) \cite{Cami_2012}. The experimental vessel is a sphere of 3 meter in diameter. An axisymmetric multicusp magnetic field (created by 36 equally spaced rings of permanent magnets with alternating polarity) confines the plasma. The field is localized near the vessel wall and a large volume of unmagnetized plasma occupies the experiment's core. An electric field applied across the multicusp field drives the edge of the plasma azimuthally. Arbitrary profiles  of azimuthal flow $v_\phi(\theta)$ can be imposed at the spherical boundary by modulating the  electric field as a function of polar angle $\theta$ using discrete electrodes. 

Results of  Ref.~\cite{Spence_2009} show that some flows generated in such a way can lead to a dynamo instability. However, in Ref.~\cite{Spence_2009} insulating boundaries are assumed, whereas in MPDX the vessel is made of aluminum whose conductivity is much higher than that of the plasma under expected experimental conditions. The goal of this paper is to generalize the results of Ref.~\cite{Spence_2009} by considering effects of varying conductivity and permeability of the  vessel on the dynamo in a model relevant to MPDX.    

To describe the plasma we use dimensionless numbers:
\begin{equation*}
M=V_0\sqrt{\frac{\rho_0}{P_0}},~Re=\frac{R_0V_0}{\nu},~Rm=\frac{R_0V_0}{\eta},~Pm=\frac{\nu}{\eta}
\end{equation*}
-- Mach, fluid Reynolds, magnetic Reynolds and magnetic Prandtl, respectively. Here $V_0$ is the peak driving velocity, $\rho_0$ and $P_0$ are the average plasma mass density and  pressure, $R_0$ is the radius of the sphere (a unit of length throughout the paper), $\nu$ and $\eta$ are the plasma kinematic viscosity and magnetic diffusivity (assumed to be constant and uniform). For given plasma parameters these numbers can be estimated from the Braginskii equations \cite{Brag_1965} (see corresponding formulas in Refs.~\cite{Spence_2009, Khalzov_2011}). Their expected values for MPDX are listed in Table \ref{t1}.  By varying temperature, density and ion species of the plasma one can change its magnetic Prandtl number by several orders of magnitude.  Such flexibility makes it possible to demonstrate  a dynamo in a laminar flow by choosing a regime with $Pm\sim1$ and $Rm\sim Re\sim10^2$. This is an advantage over the liquid metal dynamo experiments, where $Pm\sim10^{-5}$ and the flows are always turbulent.

\begin{table}[btp]
\caption{Expected parameters of MPDX}
\centering
\begin{tabular}{lccc}
\hline
\hline
Quantity & Symbol & Value & Unit \\
\hline 
Radius of sphere & $R_0$ & 1.5 & m \\
Wall thickness & $d$ & 0.05 & m \\
Peak driving velocity  & $V_0$ & $0-20$ & km/s\\
Average number density &   $n_0$  &  $10^{17}-10^{19}$ & m$^{-3}$\\
Electron temperature & $T_e$ & $2-10$ & eV \\
Ion temperature & $T_i$ &  $0.5-4$ & eV \\
Ion species &  & H, He, Ne, Ar &\\
Ion mass & $\mu_i$ & 1, 4, 20, 40 & amu\\
\hline
Mach & $M$ & $0-8$ & \\
Fluid Reynolds  & $Re$ & $0-10^5$ & \\
Magnetic Reynolds & $Rm$ & $0-2\times10^3$ & \\
Magnetic Prandtl & $Pm$ & $10^{-3}-5\times10^3$ & \\
\hline
\end{tabular}
\label{t1}
\end{table}

Our first step is to find an equilibrium velocity field capable of dynamo action. For simplicity, we do not focus on the details of plasma driving near the wall. We neglect the multicusp magnetic field and applied electric field in our consideration and assume that the velocity profile is specified at the boundary. As shown in Ref.~\cite{Khalzov_2011} for the model relevant to PCX (cylindrical prototype of MPDX), the velocity structure obtained under such assumption is the same as the velocity structure obtained with a more realistic  $\textbf{E}\times\textbf{B}$ forcing, except in a thin boundary layer.     

\begin{figure}[tbp]
\centering
\includegraphics[scale=1]{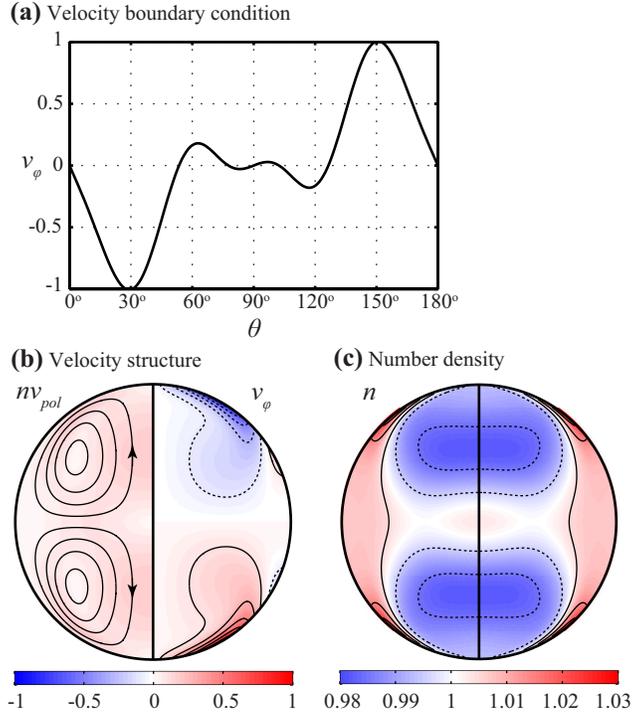}
\caption{Axisymmetric equilibrium flow of von K\'arm\'an type for Mach number $M=1$ and fluid Reynolds  number $Re=300$ used in kinematic dynamo study: (a) velocity boundary condition $v_\phi(\theta)$ adopted from Ref.~\cite{Spence_2009}; (b) structure of normalized velocity; (c) contour plot of normalized density (dashed lines denote $n<1$). Left half of (b) shows stream lines of poloidal flux $nv_{pol}$ superimposed on its absolute values depicted in colors, right half of (b) shows contour plot  of azimuthal  velocity $v_\phi$ (dashed lines denote $v_\phi<0$). Vertical lines in (b) and (c) represent the axis of symmetry.}\label{flow}
\end{figure}

The velocity field is found using the hydrodynamic part of the extended MHD code NIMROD \cite{Sovinec_2004} with an isothermal fluid model, which in non-dimensional form is
\begin{eqnarray}
\label{hd1}
\frac{\pa n}{\pa\tau}&=&-\nabla\cdot(n\v),\\
\label{hd2}
n\frac{\pa \v}{\pa\tau}&=&-n(\v\cdot\nabla)\v - \frac{\nabla n}{M^2}+ \frac{1}{Re}\left(\nabla^2\v+\frac{1}{3}\nabla(\nabla\cdot\v)\right),
\end{eqnarray}
where $\tau$, $n$ and $\v$ stand for normalized time, density and velocity, respectively: $\tau=t\,V_0/R_0$, $n=\rho/\rho_0$, $\v=\V/V_0$.  The differential plasma driving near the wall of MPDX is represented by the velocity boundary condition:
\begin{equation}\label{bc_flow}
\v\big|_{r=1}=v_\phi(\theta)\e_\phi,~~~0\leq\theta\leq\pi,
\end{equation}
where $v_\phi(\theta)$ is a function of polar angle $\theta$ with physical restriction $v_\phi(0)=v_\phi(\pi)=0$. In general, this function may be expressed as $v_\phi(\theta)=\sum\limits a_k\sin{k\theta}$, where $a_k$ are real coefficients. We use a velocity boundary condition of the von K\'arm\'an type from Ref.~\cite{Spence_2009}, shown to result in a dynamo in an incompressible flow with $Re=300$ and $Rm\gtrsim237$. It is given by $a_2=-0.4853$, $a_4=-0.5235$, $a_6=-0.0467$, $a_8=0.1516$ (Fig.~\ref{flow}a). In present study we take the Mach number $M=1$, the fluid Reynolds number $Re=300$ and the magnetic Reynolds numbers up to  $Rm=400$. These parameters can be achieved in MPDX by creating, for example, an argon plasma with $V_0=5$ km/s, $n_0=10^{18}$~m$^{-3}$, $T_e=10$ eV and $T_i=1$ eV.  

In the NIMROD simulation, we used a meshing of the poloidal plane  with 4608 quadrilateral finite elements of polynomial degree 2, and 6 Fourier harmonics in the $\phi$-direction (the azimuthal mode numbers are $0\leq m\leq5$). This resolution was sufficient for the laminar flow under consideration. We took a non-moving fluid ($\v=0$) with uniform density ($n=1$) as the initial state and evolved Eqs.~(\ref{hd1}), (\ref{hd2}) with the boundary condition given by Eq.~(\ref{bc_flow})  until a steady state was reached. The resulting  flow for $Re=300$ and $M=1$ is shown in Fig.~\ref{flow}. The velocity field $\v(r,\theta)$ is axisymmetric and hydrodynamically stable with respect to perturbations with $m>0$.

The main results of the paper are obtained by solving the kinematic dynamo problem with this velocity field,
\begin{equation}\label{kin}
\gamma\B=Rm\nabla\times(\v\times\B)+\nabla^2\B,~~~\nabla\cdot\B=0,
\end{equation}
for unknown magnetic field $\B$ and normalized dynamo growth rate $\gamma=\Gamma R_0^2/\eta$. We represent the divergence-free field as an expansion in a spherical harmonic basis~\cite{Bullard_1954}:
\begin{eqnarray}
\label{Br} B_r&=&\sum\limits_{l=m}^{L}\frac{l(l+1)S_lY_l^m}{r^2}\,e^{im\phi},\\
\label{Bt} B_\theta&=&\sum\limits_{l=m}^{L}\bigg[\frac{1}{r}\frac{\pa S_l}{\pa r}\frac{\pa Y_l^m}{\pa\theta}+\frac{imT_lY_l^m}{r\sin{\theta}}\bigg]\,e^{im\phi},\\
\label{Bp} B_\phi&=&\sum\limits_{l=m}^{L}\bigg[\frac{imY_l^m}{r\sin{\theta}}\frac{\pa S_l}{\pa r}-\frac{T_l}{r}\frac{\pa Y_l^m}{\pa\theta}\bigg]\,e^{im\phi},
\end{eqnarray} 
where $S_l(r)$ and $T_l(r)$ are functions of $r$ only and $Y_l^m(\theta)$ are spherical harmonics related to the associated Legendre polynomials by  $Y_l^m(\theta)=P_l^m(\cos{\theta})$.  Since the velocity  is axisymmetric, we consider  each azimuthal mode $m$ separately. The summation in Eqs.~(\ref{Br})-(\ref{Bp}) is truncated at some spherical harmonic  $L$ ($L=20$ provides a satisfactory convergence in these studies). Substituting Eqs.~(\ref{Br})-(\ref{Bp}) into Eq.~(\ref{kin}) and using the orthogonal properties of spherical harmonics, one obtains for $m\leq l\leq L$: 
\begin{eqnarray}
\label{S} \gamma S_l&=&\frac{\pa^2 S_l}{\pa r^2}-\frac{l(l+1)S_l}{r^2}+Rm\,A_l^m\sum\limits_{j=m}^{L}\bigg[I^{(1)}_{lj}S_j\nonumber\\ 
&-& I^{(2)}_{lj}\frac{\pa S_j}{\pa r} + I^{(3)}_{lj}T_j \bigg],\\
\label{T} \gamma T_l&=&\frac{\pa^2 T_l}{\pa r^2}-\frac{l(l+1)T_l}{r^2}-Rm\,A_l^m\sum\limits_{j=m}^{L}\bigg[\overline{I^{(1)}_{jl}}T_j\nonumber\\ 
&+& \frac{\pa}{\pa r}\left(I^{(2)}_{lj}T_j+I^{(3)}_{lj}\frac{\pa S_j}{\pa r} + I^{(4)}_{lj}S_j\right)+ \overline{I^{(4)}_{jl}}\frac{\pa S_j}{\pa r} \bigg].
\end{eqnarray} 
Here the bar above a symbol denotes its complex conjugate, $A_l^m$ is a numerical factor
$$
A_l^m=\frac{(2l+1)(l-m)!}{2l(l+1)(l+m)!},
$$
and $I^{(1-4)}_{lj}(r)$ are functions of $r$ given by the integrals:
\begin{eqnarray}
\label{I1} I^{(1)}_{lj}&=&\frac{j(j+1)}{r}\int\limits_0^\pi Y_j^m\bigg[v_\theta \frac{\pa Y_l^m}{\pa\theta}\sin{\theta}-imv_\phi Y_l^m\bigg]d\theta,\\
\label{I2} I^{(2)}_{lj}&=&\int\limits_0^\pi v_r\bigg[\frac{\pa Y_l^m}{\pa\theta}\, \frac{\pa Y_j^m}{\pa\theta}\,\sin{\theta}+\frac{m^2Y_l^m Y_j^m}{\sin{\theta}} \bigg]d\theta,\\
\label{I3} I^{(3)}_{lj}&=&im\int\limits_0^\pi \frac{\pa v_r}{\pa\theta}\,Y_l^mY_j^m\,d\theta,\\
\label{I4} I^{(4)}_{lj}&=&\frac{j(j+1)}{r}\int\limits_0^\pi Y_j^m\bigg[v_\phi\frac{\pa Y_l^m}{\pa\theta}\sin{\theta} + imv_\theta Y_l^m\bigg]d\theta.
\end{eqnarray} 
Note that Eqs.~(\ref{S})-(\ref{I4}) are valid for any axisymmetric velocity field. To calculate the integrals in Eqs.~(\ref{I1})-(\ref{I4}), we interpolate the velocity field on a uniform polar grid (typically with $N_r=50$ radial and $N_\theta=1000$ angle grid points) and  use the trapezoidal rule of integration. 

Eqs.~(\ref{S}), (\ref{T}) should be supplemented with boundary conditions for functions $S_l(r)$ and $T_l(r)$. The absence of a singularity in the field at the center of the sphere requires 
\begin{equation}\label{BC0} 
S_l\big|_{r=0}=0,~~~T_l\big|_{r=0}=0.
\end{equation}
The outer boundary conditions depend on the properties of the shell. To avoid undesired diversion of flow-driving current into the shell, the inner surface in MPDX is covered with an insulating coating (Fig.~\ref{MPDX_fig}c). Thus, the normal component of current is zero at $r=1$, i.e.,
\begin{equation}\label{BCT} 
T_l\big|_{r=1}=0.
\end{equation}
To derive the condition for $S_l$ at $r=1$, we consider the model shown in Fig.~\ref{MPDX_fig}c and use the general boundary conditions for normal and tangential components of the magnetic field at the interface between two media with different relative magnetic permeabilities  $\mu_1$ and $\mu_2$:
$$
B_{1n}=B_{2n},~~~\frac{B_{1t}}{\mu_1}=\frac{B_{2t}}{\mu_2}.
$$
We also assume that the insulating  coating is thin enough that it has no impact on profile of $S_l$. Then the resulting equations are (omitting ``$l$" in $S_l$)
\begin{eqnarray}
\label{pw}r=1&:&S=S_w,~~\frac{1}{\mu}\frac{\pa S}{\pa r}=\frac{1}{\mu_w}\frac{\pa S_w}{\pa r},\\
\label{ww}1<r<1+\frac{d}{R_0}&:&\frac{\eta}{\eta_w}\gamma S_w=\frac{\pa^2 S_w}{\pa r^2}-\frac{l(l+1)S_w}{r^2},\\
\label{wv}r=1+\frac{d}{R_0}&:&S_w=S_v,~~\frac{1}{\mu_w}\frac{\pa S_w}{\pa r}=\frac{\pa S_v}{\pa r},\\
\label{vv}r>1+\frac{d}{R_0}&:&S_v\propto r^{-l},
\end{eqnarray}
where Eq.~(\ref{ww}) is derived for a stationary wall with thickness $d$, symbols with subscripts refer to wall (``$w$") and vacuum (``$v$"), and symbols without subscript refer to plasma. We assume that $d$ is small, so that the variations of $S_l$ in the wall are small too. This is the thin-wall approximation \cite{Roberts_2010}, it applies if  $d\ll R_0$ and $d\ll|\eta_w/\Gamma|^{1/2}$. Under these assumptions, Eqs.~(\ref{pw})-(\ref{vv}) are reduced to
\begin{equation}
\label{BCS} \left(\frac{\pa S_l}{\pa r}\,(1+lc_\mu)+S\,(l\mu+\gamma c_\sigma)\right)\bigg|_{r=1}=0,
\end{equation}
where we have used the relation $\eta=c^2/(4\pi\sigma\mu)$ between magnetic diffusivity $\eta$ and electric conductivity $\sigma$ of a medium ($c$ is the speed of light), and introduced the wall conductivity parameter $c_\sigma$ and the wall permeability parameter $c_\mu$,
\begin{equation}
\label{c} c_\sigma=\frac{\sigma_w d}{\sigma R_0},~~~c_\mu=\frac{\mu_w d}{R_0}.
\end{equation}
Eq.~(\ref{BCS}) is obtained for a stationary wall without requiring the no-slip boundary condition for plasma velocity,  in contrast to analogous equation (14a) from Ref.~\cite{Roberts_2010}.

\begin{figure}[h]
\centering
\includegraphics[scale=1]{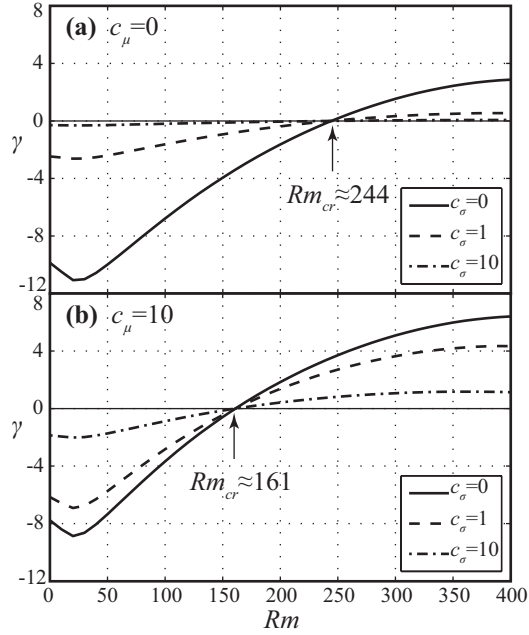}
\caption{Dependence of dynamo growth rate $\gamma$ on magnetic Reynolds number $Rm$ for different values of the wall parameters $c_\sigma$ and $c_\mu$.}\label{cmu}
\end{figure} 
\begin{figure}[h]
\centering
\includegraphics[scale=1]{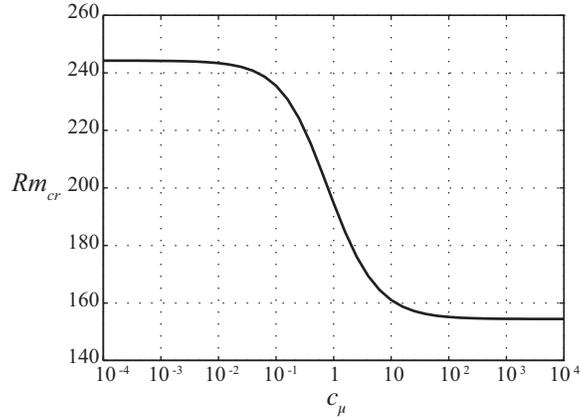}
\caption{Dependence of critical magnetic Reynolds number $Rm_{cr}$ on the wall permeability parameter $c_\mu$.}\label{Rmcr}
\end{figure}

Eqs.~(\ref{S}), (\ref{T}), (\ref{BC0}), (\ref{BCT}), (\ref{BCS}) constitute an eigenvalue problem for the dynamo growth rate $\gamma$ and unknown eigen-functions $S_l$ and $T_l$. In order to solve it, we apply the finite difference method to Eqs.~(\ref{S}), (\ref{T})  and discretize $S_l$ and $T_l$ for each harmonic $l$ ($m\leq l\leq L$) on  a uniform grid at $0\leq r\leq1$ with $N_r$ equal intervals.  Eqs.~(\ref{BC0}) and (\ref{BCT}) are straightforward to implement in the finite difference scheme, while  Eq.~(\ref{BCS}) is taken into account by using an extra  (ghost) grid point to approximate the derivative. The resulting system of  $(L-m+1)(2N_r-1)$ linear algebraic equations for $(L-m+1)(2N_r-1)$ unknowns is cast in the form of a matrix eigenvalue equation, which is solved in MATLAB. The developed scheme has been successfully benchmarked against the results of the kinematic dynamo study from Ref.~\cite{Dudley_1989}.

Here we report  the results of solving the kinematic dynamo eigenvalue problem [Eqs.~(\ref{S}), (\ref{T}), (\ref{BC0}), (\ref{BCT}), (\ref{BCS})] with the velocity shown in Fig.~\ref{flow}, and for the relative permeability of the plasma $\mu=1$, magnetic  Reynolds numbers $Rm=0-400$ and varying wall parameters $c_\sigma$ and $c_\mu$. We consider only the most unstable (or least decaying) $m=1$ azimuthal mode. The convergence of the numerical scheme is checked by comparing simulations at different resolutions. Results reported here are obtained by using a maximum number of spherical harmonics $L=20$ and number of radial grid points $N_r=50$. 

The results are summarized in Figs.~\ref{cmu} and \ref{Rmcr}. In the present case $\gamma$ is always real, so the dynamo threshold  $Rm_{cr}$  corresponds to the condition  $\gamma=0$.   As seen in Fig.~\ref{cmu}, $Rm_{cr}$ required for the onset of the dynamo does not depend on the wall conductivity parameter $c_\sigma$. This is because $c_\sigma$ drops out of the problem when $\gamma=0$, as follows from Eq.~(\ref{BCS}).  However,  $c_\sigma$ affects the dynamo growth rate: larger values of $c_\sigma$ (larger wall conductivity) lead to lower $|\gamma|$. Therefore, in the limit of a perfectly conducting shell  no growing field is possible, since $\gamma\to0$. 

The wall permeability parameter $c_\mu$ has a strong influence on both dynamo threshold $Rm_{cr}$ and growth rate $\gamma$.  A ferritic wall facilitates dynamo action.  As shown in Fig.~\ref{Rmcr}, the critical magnetic Reynolds number $Rm_{cr}$  decreases with increase of $c_\mu$: from $Rm_{cr}\approx244$ when $c_\mu=0$ to $Rm_{cr}\approx154$ when $c_\mu\to\infty$. These results are consistent with previous theoretical dynamo studies in other geometries \cite{Avalos_2003, Avalos_2005, Gissinger_2008, Gissinger_2009}, which indicated reduction of $Rm_{cr}$ for the ferritic-wall boundary conditions.

Estimates for typical parameters of MPDX show that its wall is very conducting and non-ferritic with \mbox{$c_\sigma\approx30$} and \mbox{$c_\mu\approx0$}. Under these conditions,  dynamo action is achievable for the considered flow if \mbox{$Rm\gtrsim244$}, the respective dynamo growth rate  at \mbox{$Rm=400$} is \mbox{$\Gamma\approx3.6$~s$^{-1}$}.

In summary, we have studied the influence of finite  conductivity and  permeability of the wall on a plasma dynamo in a sphere. Our results show that the dynamo threshold is affected only by the wall permeability, while the dynamo growth rate depends on both wall properties. 

The authors  wish to thank C. Sovinec for valuable help and discussions related to \mbox{NIMROD}.

\end{document}